\documentclass[twocolumn,pra,showpacs,amsmath,amssymb]{revtex4}
\usepackage{graphicx}
\usepackage{dcolumn}
\usepackage{bm}

\begin{document}

\title{Quantum computing based on space states without charge transfer}

\author{S. Filippov$^{1}$}
\author{V. Vyurkov$^{1}$}
\email{vyurkov@ftian.ru}
\author{L. Gorelik$^{2}$}
\email{gorelik@chalmers.se}

\affiliation{$^1$ Institute of Physics and Technology of the
Russian Academy of Sciences, Moscow,
Russia\\
$^2$ Dept. Applied Physics, Chalmers University of
Technology and Goteborg University, Goteborg, Sweden}

\begin{abstract}
An implementation of a quantum computer based on space states in
double quantum dots is discussed. There is no charge transfer in
qubits during calculation, therefore, uncontrollable entanglement
between them due to long-range Coulomb interaction is suppressed.
Other plausible sources of decoherence caused by interaction with
phonons and gates could be substantially suppressed in the
structure too. We also demonstrate how all necessary quantum logic
operations, initialization, writing, and read-out could be carried
out in the computer.
\end{abstract}


\maketitle

\section{\label{introduction}Introduction}

Almost all recent experimental realizations of quantum computation
accomplished on several solid state qubits and even theoretical
proposals of new devices were based on spin states. Much cited
papers of Kane \cite{kane} and DiVincenzo et al.
\cite{loss98,burkard99,vrijen} just concerned nucleus or electron
spin encoding in solid state implementation of a quantum computer
(QC). The attractiveness of spin states in solid state QC was
mainly caused by quite long decoherence time: hours for a nucleus
spin and milliseconds for electron spin seem attainable
\cite{kane,vrijen}.

Nevertheless, space or charge states for quantum encoding do not
seem less prospective although two main disadvantages were
commonly mentioned \cite{burkard99}. Firstly, much less
decoherence time was expected for charge states. Secondly, charge
transfer results in uncontrollable interaction between even
distant qubits due to long-range Coulomb forces. However, owing to
the pioneer papers
\cite{fedichkin-yanchenko-valiev,fedichkin-yanchenko-valiev::nanotechnology,fedichkinPRA}
now it is clear that for fairly small energy gap between different
qubit states decoherence caused by phonons might be very weak.
This has opened trends in quantum computers based on double
quantum dots. Recently they were intensively studied in
experiments \cite{fujisawa,gorman05,shinkai}. The similar
structures based on double donor atoms in silicon were proposed in
Ref. \cite{hollenberg-charge04}.

Here we withdraw the second shortcoming of a space state based QC
discussing a construction \cite{vyurkov00,vyurkov-2}, where no
charge transfer occurs during computation. Besides, the
computation may be much faster than that in a spin based QC.
Moreover, the read-out of charge states is looking easier and
faster than that of spin states.

\section{\label{section::qubit:operation}Qubit and its operation}

\begin{figure}
\includegraphics{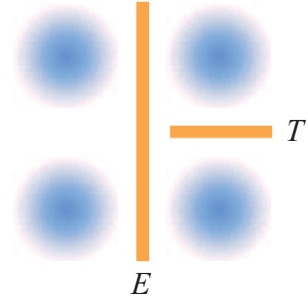}
\caption{\label{figure:qubit:structure} Qubit structure.}
\end{figure}

A qubit is implemented in two double quantum dots (DQD) (Fig.
\ref{figure:qubit:structure}). A DQD consists of a pair of quantum
dots with a single electron. The electrode   operates on the
strength of exchange interaction between electrons in DQDs. The
electrode varies tunneling coupling between quantum dots
constituting a DQD.

\begin{figure}
\includegraphics{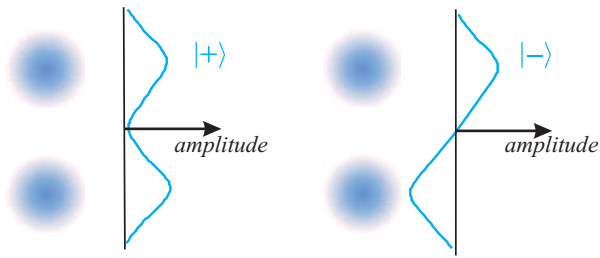}
\caption{\label{figure:wave:functions} Wave functions of an
electron in a DQD (two lowest states).}
\end{figure}

We designate two lowest states of an electron in a DQD as
$|+\rangle$ and $|-\rangle$. The state

\begin{equation}
|+\rangle = \frac{1}{\sqrt{2}}\Big( \psi(r-r_{1}) +
\psi(r-r_{2})\Big)
\end{equation}

\noindent is symmetric and the state

\begin{equation}
|-\rangle = \frac{1}{\sqrt{2}}\Big( \psi(r-r_{1}) -
\psi(r-r_{2})\Big)
\end{equation}

\noindent is anti-symmetric (Fig. \ref{figure:wave:functions}).
Here $r_{1}$ and $r_{2}$ are coordinates of the center of the
first and second quantum dots, $\psi(r)$ is an electron wave
function in a dot, it decays as $e^{-r/a}$ outside the dot, the
magnitude $a$ of depends on a potential barrier height which may
be controlled by a voltage on gate $T$.

Two basic computational states of the qubit composed of two DQDs
are

\begin{equation}
\label{0:1:simple} |0\rangle = |+-\rangle, \qquad |1\rangle =
|-+\rangle.
\end{equation}

In these states an electron in the first DQD is in symmetric state
and another electron in the second DQD is in anti-symmetric state,
and vice versa. As an exchange interaction is to be involved
later, one should take into account the Fermi-Dirac statistics of
electrons. When two electrons are in the triplet spin state (a
total spin $S = 1$), i.e., have a symmetric spin configuration,
their space wave function is anti-symmetric with respect to
permutation. Hereafter we assume an overall spin polarization of
electrons in the system and, therefore, instead of states
(\ref{0:1:simple}) one should use the following states:

\begin{eqnarray}
\label{0:comlex}{\left| {0} \right\rangle}  = {\frac{{1}}{{\sqrt
{2}}} }\Big( {{\left| { + _{1} - _{2}} \right\rangle}  -
{\left| { + _{2} - _{1}}  \right\rangle}} \Big),\\
\label{1:complex} {\left| {1} \right\rangle}  = {\frac{{1}}{{\sqrt
{2}}} }\Big( {{\left| { - _{1} + _{2}}  \right\rangle}  - {\left|
{ - _{2} + _{1}}  \right\rangle}} \Big),
\end{eqnarray}

\noindent where the indices $1$ and $2$ correspond to the first
and second electron, respectively. Any superpositional state of a
qubit

\begin{equation}
\label{arbitrary:qubit:state}
\frac{a|0\rangle+b|1\rangle}{\sqrt{a^2+b^2}}
\end{equation}

\noindent is created with the help of gate electrodes. Here $a$
and $b$ are arbitrary complex numbers. Thus qubit states belong to
the Hilbert sub-space produced by orthogonal states $|+-\rangle$
and $|-+\rangle$.

Worth noting no charge transfer is required to produce any of the
states (\ref{arbitrary:qubit:state}). It can be easily confirmed
that for any numbers $a$ and $b$ the probability to find an
electron in any quantum dot is always equal to 0.5. This is just a
reason why a qubit based on a pair of DQDs is required instead of
a qubit based on a single DQD.

Hereafter, the unitary transformations of a qubit are described by
the Hamiltonian in matrix presentation in basis of states
(\ref{0:comlex}) and (\ref{1:complex})

\begin{equation}
\mathcal{H} = A \left(%
\begin{array}{cc}
  0 & 1 \\
  1 & 0 \\
\end{array}%
\right) + P \left(%
\begin{array}{cc}
  1 & 0 \\
  0 & -1 \\
\end{array}%
\right)
\end{equation}

\noindent where the factors $A$ and $P$ depend upon a voltage
applied to gates $E$ and $T$, respectively. The associated unitary
time evolution is determined by the operator

\begin{equation}
\hat{U}(t) = \hat{T} \exp\left[ -\frac{i}{\hbar} \int _{0}^{t}
\mathcal{H}(\tau) d \tau\right],
\end{equation}

\noindent where $\hat{T}$ is a time ordering operator.

The amplitude shift can be realized by applying voltage to
electrode $E$ which operates upon the strength of exchange
interaction between electrons located in surrounding DQDs (Fig.
\ref{figure:qubit:structure}). The applied positive voltage
diminishes the potential barrier height and augments the wave
function overlap. It leads to enhancement of exchange interaction.
In particular, an amplitude flip resulting in a transition of a
state $|0\rangle$ into a state $|1\rangle$ can be performed by
means of appropriate pulse amplitude and duration. An exchange
interaction is based on Coulomb interaction between electrons

\begin{equation}
\hat{U}_{C}(r_1,r_2) = \frac{e^2}{\kappa |r_1 - r_2|},
\end{equation}

\noindent where $r_1$ and $r_2$ are respectively the coordinates
of first and second electron, $\kappa$ is a permittivity of an
environment. The direct calculation of matrix elements of operator
$\hat{U}_{C}$ for transitions from basic qubit states $|+-\rangle$
and $|-+\rangle$ to complimentary states $|++\rangle$ or
$|--\rangle$ gives zero in contrast with transitions between basic
states. Obviously, beforehand the states $|++\rangle$ and
$|--\rangle$ should be also presented in the form like
(\ref{0:comlex})-(\ref{1:complex}). Moreover, the transitions to
these states are prohibited by symmetry conservation. Indeed,
$|++\rangle$ and $|--\rangle$ states have additional symmetry with
respect to inversion over $E$-electrode axis. The Coulomb operator
$\hat{U}_{C}$ can not break this symmetry.

A phase shift of a qubit is operated by the gate electrode $T$
acting on a tunneling coupling between the associated quantum
dots. When a positive potential is applied tunneling between
constituent quantum dots is reinforced and the energy difference
$\Delta \varepsilon$ between $|+\rangle$ and $|-\rangle$ states of
this DQD becomes greater. It results in steadily rising phase
difference between $|+\rangle$ and $|-\rangle$ states. In
particular, for a proper impulse duration and amplitude the phase
difference achieves $\pi$, i.e., a phase flip occurs. It should be
outlined that no tunneling can happen during a phase shift
operation as it needs a definite bias between quantum dots (see
for the section "Read-out").

\section{\label{two:qubit:operations} Two-qubit operations}

\begin{figure}
\includegraphics{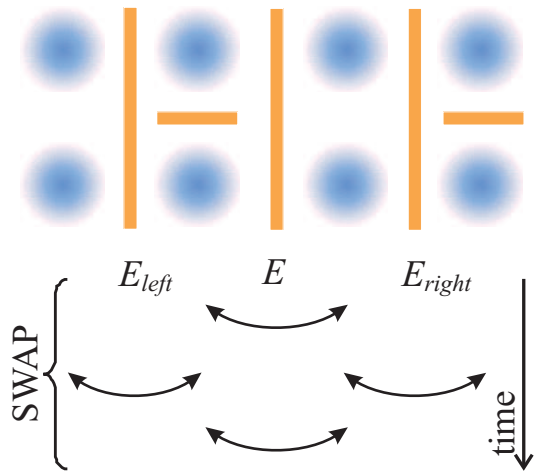}
\caption{\label{figure:SWAP} SWAP realization.}
\end{figure}

The electrodes of $E$-type placed between adjacent qubits make
possible to perform two-qubit operations. The most simple for
realization is a SWAP operation, that is, an exchange of states
between neighbor qubits. This operation is fulfilled merely by
sequential application of voltage pulses to electrodes $E$
resulting in exchange of states between surrounding DQDs. Thus,
the SWAP operation between neighbor qubits can be easily
represented by the formula $SWAP=\hat{E} \cdot NOT_1 \cdot NOT_2
\cdot \hat{E}$, where $\hat{E}$ is exchange operation with the
help of $E$-electrode located between the first and the second
qubits, $NOT_1$ and $NOT_2$ are NOT operations in the first and in
the second qubits respectively (Fig. \ref{figure:SWAP}). The SWAP
operation allows to move any qubit along a chain and thus put in
contact any pair of qubits.

Realization of CNOT is based on the square root of the SWAP
operator, which is given by the expression $SWAP=\hat{E} \cdot
\sqrt{NOT_1} \cdot \sqrt{NOT_2} \cdot \hat{E}$. The definite
performance of CNOT operation is presented in Appendix. Other
2-qubit logic operations could be also composed in a similar way.
The rate of 1-qubit and 2-qubit operations depend on the strength
of exchange interaction and tunneling coupling augmented by
voltage applied to electrodes. The offered construction permits,
in principle, a clock speed up to 1THz compared to 1 GHz for
electron spin based QC \cite{vrijen} or 75 kHz for nucleus spin
based QC \cite{kane}.

\section{Initialization of a computer}

One way of initial state creation in the QC is to input electrons
in a necessary state through the end of preliminary empty qubit
chain. An electron can be supplied to the extreme DQD by some
single-electron device, for instance, a turnstile. A symmetric
state $|+\rangle$ can be prepared by applying a fairly high
voltage to an electrode $T$ to make a sufficient energy difference
between $|+\rangle$ and $|-\rangle$ states. The first one is a
ground state of an electron in a DQD. Thus for fairly large energy
difference between ground $|+\rangle$ and excited $|-\rangle$
states with respect to a thermal energy $kT$ the occupancy of the
upper state can be done negligible. A state $|+\rangle$ can be
easily inverted into a state $|+\rangle$ (if required) when a DQD
is biased by means of some additional electrode placed near one
quantum dot. Afterwards, an electron with a formed state can be
pushed from the starting DQD to any DQD in a chain by voltage
pulses successively applied to $E$-electrodes along the path. In
this way one can pump electrons one by one along a qubit chain and
fill all DQDs.

Another possibility of initialization implies the formation of
necessary state immediately in a particular qubit by means of
foregoing procedure.

All electrons in the proposed QC should be spin-polarized as an
exchange interaction depends on a spin configuration. Two
opportunities seem like plausible. Obviously, it could be done by
external magnetic field and cooling. Another way is to supply
electrons from a ferromagnetic contact.

There is a possibility to speed up $|+\rangle$ or $|-\rangle$
states creation even under condition $\Delta \varepsilon < kT$
when a thermal relaxation is evidently ineffective. The procedure
merely looks like the reversed read-out considered in the next
section.

\section{Read-out}

To read out the information accumulated in the resultant register
of a QC one should measure the state of DQDs. The following
procedure is proposed. The voltage applied to the gate $T$ makes
an essential energy gap $\Delta \varepsilon$ between $|+\rangle$
and $|-\rangle$ states. The mean energy of a system when electron
occupies one quantum dot is situated just in the middle of the
gap. When this DQD is biased by a voltage $\frac{\Delta
\varepsilon}{2e}$ by means of outer electrode placed near one
quantum dot resonant tunneling of an electron to a quantum dot
occurs. To what dot of two depends on initial state. Surely, the
state when an electron is located in one quantum dot is not an
eigen state of a system. Actually on applying a bias the electron
begins to oscillate between quantum dots. But what dot it visits
first depends on the initial state of a DQD, whether it is
$|+\rangle$ or $|-\rangle$. One could switch off $T$-voltage,
i.e., tunnelling, right at the moment when electron is located in
one quantum dot. Measuring a conductance of a quantum wire placed
nearby one can easily distinguish what quantum dot contains the
electron. Moreover, a current-voltage curve of a single wire
placed along a resultant qubit chain could, in principle, provide
the whole information about its charge state \cite{tanamoto99}.

\section{Decoherence: interaction with gates and phonons}

The weaker is decoherence the longer is a fault tolerant time of a
QC. The decoherence in our QC circuit arises when a state of some
DQD is altered from $|+\rangle$ to $|-\rangle$, or vice versa. Two
comprehensible sources of decoherence in the system are voltage
fluctuations and phonons. One more source mentioned recently is
the Joule loss in a nearby metallic electrode caused by
displacement of charge inside a qubit
\cite{vyurkov::image-force,woodford07}. This kind of decoherence
is absent in the QC under consideration because no charge transfer
occurs.

Voltage fluctuations were discussed in \cite{kane}. There resonant
transitions between states were induced by associated spectrum
component of a voltage noise. However, a more dangerous mechanism
is a phase diffusion of a qubit. It takes place when voltage
fluctuations shift phase of a qubit slightly in random way. In
some time that leads to a definite phase shift. In the present
construction voltage fluctuations on electrodes $T$ and $E$ can
not affect transitions between $|+\rangle$ and $|-\rangle$ states
owing to symmetry of the structure.  One more advantage relevant
to offered implementation is that manipulation with some qubit
does not perturbs the state of neighbor ones.

However, optical gates to control a potential barrier transparency
\cite{openov99,sanders99,fedirko00} could be used instead of
voltage gates. Extremely strong dependence of a photo-stimulated
tunneling on a beam polarization permits a precise addressing to a
definite qubit \cite{fedirko00}.

As for the decoherence induced by phonons, the cheerful results
were obtained in
\cite{fedichkin-yanchenko-valiev,fedichkin-yanchenko-valiev::nanotechnology,fedichkinPRA}.
They much contradict with intuitive expectations. Here we restrict
the discussion by rather qualitative consideration. The states
$|+\rangle$ and $|-\rangle$ have very small energy difference
$\Delta \varepsilon$ in an idle qubit. The same is the energy of
an acoustic phonon $\varepsilon$ required to enforce a transition
between these states. The lower is the energy the smaller is the
matrix element of these transitions. One reason is a long wave
length of a relevant phonon. In
\cite{fedichkin-yanchenko-valiev::nanotechnology} a probability
$\tau^{-1}$ of spontaneous emission of acoustic phonons in DQD was
calculated. There were obtained dependencies $\tau \sim \Delta
\varepsilon^{-5}$ for deformation acoustic phonons and $\tau \sim
\Delta \varepsilon^{-3}$ for piezoelectric acoustic phonons. The
latter do not exist in $Si$ but dominate in compounds. When two
$GaAs/AlGaAs$ quantum dots constituting a DQD are separated by a
distance $r=22\ nm$ the values $\tau \sim 10^{-2}\ s$ for
piezoelectric acoustic phonons and $\tau \sim 10^{-6}\ s$ for
deformation acoustic phonons are attainable. The stimulated
emission and absorption, and multi-phonon processes were beyond
the frame of
\cite{fedichkin-yanchenko-valiev,fedichkin-yanchenko-valiev::nanotechnology,fedichkinPRA}.
However, they may be significant when $\Delta \varepsilon < kT$.
To roughly evaluate the stimulated emission one should multiply
the above magnitude of $\tau$ by a small factor $\frac{\Delta
\varepsilon}{kT}$ originating from the Bose-Einstein statistics
for $\Delta \varepsilon \ll kT$, $n(\Delta\varepsilon) = [
e^{\Delta\varepsilon / kT} - 1 ]^{-1} \approx \frac{kT}{\Delta
\varepsilon }$. Nevertheless, the probability to emit or to absorb
a phonon still drops with energy difference $\Delta \varepsilon$.
It looks amazing as if one could sustain the coherence between
states separated by energy less than $kT$ for an arbitrary long
time.

In our opinion, in reality two-phonon processes set a limit for
decoherence time. During transition from a state $|+\rangle$ to a
state $|-\rangle$ one phonon is emitted and another absorbed. The
energy of phonons involved is about $kT$, that is, independent on
energy split $\Delta \varepsilon$ between states $|+\rangle$ and
$|-\rangle$.

Here we calculate the probability of transitions from state
$|+-\rangle$ to states $|-+\rangle$, $|++\rangle$, or $|--\rangle$
due to absorption of a phonon with a wave vector ${\bf q}$ and
emission of a phonon ${\bf q}'$ in the second order of the
perturbation theory. The probability of the transition from the
state $|+-\rangle$ to an arbitrary final state $|f\rangle$ is

\begin{widetext}
\begin{eqnarray}
 W({|+ -\rangle} \to {|f\rangle}) = \frac{2\pi}{\hbar}
\sum\limits_{{\bf q}} \sum\limits_{{\bf q}'} && \vert F_{{\bf q}}
\vert ^{2} \vert F_{{\bf q}'} \vert ^{2} \left( {n (\hbar \omega
_{{\bf q}}) + 1} \right) n(\hbar \omega_{{\bf q}'})
\nonumber \\
&& \times \frac{\left| \sum\limits_{z} \langle f | e^{i {\bf
q}'{\bf r}} |z\rangle \langle z | e^{-i {\bf q}{\bf r}} |+-\rangle
\right|^{2}} {\left( \varepsilon_{z} - \varepsilon_{+-} - \hbar
\omega _{{\bf q}} \right)^{2}} \delta (\varepsilon _{f} + \hbar
\omega _{{\bf q}'} - \varepsilon _{+-} - \hbar \omega_{{\bf q}} ).
\end{eqnarray}
\end{widetext}

Here $F_{{\bf q}}$ is electron-phonon coupling, $\omega_{{\bf
q}}$, $\omega_{{\bf q}'}$ are phonons frequencies,
$\varepsilon_{+-}$ and $\varepsilon_{f}$ are energies of initial
and final states, the summation over intermediate states
$|z\rangle$ is made. Because of a high energy of upper states only
the states $|+-\rangle$, $|-+\rangle$, $|++\rangle$, and
$|--\rangle$ can be put into account as intermediate ones. As far
as phonons with energy $\hbar \omega_{{\bf q}} \approx kT \gg
\Delta \varepsilon$ mostly contribute to the probability the
denominator $\left( \varepsilon_{z} - \varepsilon_{+-} - \hbar
\omega _{{\bf q}} \right)^{2}$ can be reduced to $(kT)^{2}$. Then
the evaluation of probability gives the dependence $W \sim T^{6}$
for deformation acoustic phonons and the dependence $W \sim T^{2}$
for piezoelectric acoustic phonons. Hence, just two-phonon
processes set a limit of decoherence rate when the energy
separation  is decreasing to zero.

\section{Conclusions}
In conclusion, an implementation of a space state based quantum
computer without charge transfer is discussed. A qubit consists of
four quantum dots with two electrons. Evolution of the system is
controlled with gate voltages operating on tunneling coupling and
strength of exchange interaction. As there is no charge transfer
during calculation, therefore, uncontrollable entanglement between
qubits due to long-range Coulomb forces is suppressed. The
decoherence caused by metallic electrodes is much diminished in
the structure. Phonon-induced decoherence is limited by two-phonon
processes. High-speed computation and long fault tolerant time
look as feasible.

\begin{acknowledgments}
The research was supported by NIX Computer Company
(science@nix.ru), grant $\#$ F793/8-05, and also via the grant of
The Royal Swedish Academy of Sciences.
\end{acknowledgments}

\appendix*
\section{Two-qubit operations}


Here we consider 2-qubit operations in detail. The initial state
of pair of qubits is a linear combination of the following basic
states:

\begin{equation}
\left. \begin{array}{cc}
{\rm 1^{st}qubit} & {\rm 2^{nd}qubit}\\
 +- & +- \\
 +- & -+ \\
 -+ & +- \\
 -+ & -+ \\
\end{array}\right| \begin{array}{c}
  {\rm basis\ state} \\
  (1000)^{T} \\
  (0100)^{T} \\
  (0010)^{T} \\
  (0001)^{T} \\
\end{array}
\end{equation}

Unfortunately, this basis does not cover all states originating
during SWAP operation. Indeed, if we apply voltage to electrode
placed between qubits in question, then   state will become
which does not belong to the original sub-space. Thus, the
sub-space should be extended as follows:

\begin{equation}
\left. \begin{array}{cc}
{\rm 1^{st}qubit} & {\rm 2^{nd}qubit}\\
 +- & +- \\
 ++ & -- \\
 +- & -+ \\
 -+ & +- \\
 -- & ++ \\
 -+ & -+ \\
\end{array}\right| \begin{array}{c}
  {\rm basis\ state} \\
  (100000)^{T} \\
  (010000)^{T} \\
  (001000)^{T} \\
  (000100)^{T} \\
  (000010)^{T} \\
  (000001)^{T} \\
\end{array}
\end{equation}

Just this requirement makes the realization of CNOT in the
structure under consideration non-trivial. Hereafter, we represent
all operations by matrices. The NOT operations in the first and in
the second qubit are

\begin{eqnarray}
 NOT_{1} = \left( {{\begin{array}{*{20}c}
 {0} \hfill & {0} \hfill & {0} \hfill & {1} \hfill & {0} \hfill & {0} \hfill
\\
 {0} \hfill & {1} \hfill & {0} \hfill & {0} \hfill & {0} \hfill & {0} \hfill
\\
 {0} \hfill & {0} \hfill & {0} \hfill & {0} \hfill & {0} \hfill & {1} \hfill
\\
 {1} \hfill & {0} \hfill & {0} \hfill & {0} \hfill & {0} \hfill & {0} \hfill
\\
 {0} \hfill & {0} \hfill & {0} \hfill & {0} \hfill & {1} \hfill & {0} \hfill
\\
 {0} \hfill & {0} \hfill & {1} \hfill & {0} \hfill & {0} \hfill & {0} \hfill
\\
\end{array}}}  \right),
\\
NOT_{2} = \left( {{\begin{array}{*{20}c}
 {0} \hfill & {0} \hfill & {1} \hfill & {0} \hfill & {0} \hfill & {0} \hfill
\\
 {0} \hfill & {1} \hfill & {0} \hfill & {0} \hfill & {0} \hfill & {0} \hfill
\\
 {1} \hfill & {0} \hfill & {0} \hfill & {0} \hfill & {0} \hfill & {0} \hfill
\\
 {0} \hfill & {0} \hfill & {0} \hfill & {0} \hfill & {0} \hfill & {1} \hfill
\\
 {0} \hfill & {0} \hfill & {0} \hfill & {0} \hfill & {1} \hfill & {0} \hfill
\\
 {0} \hfill & {0} \hfill & {0} \hfill & {1} \hfill & {0} \hfill & {0} \hfill
\\
\end{array}}}  \right).
\end{eqnarray}


The SWAP operation between adjacent DQDs in neighbor qubits is

\begin{equation}
\hat {E} = \left( {{\begin{array}{*{20}c}
 {0} \hfill & {1} \hfill & {0} \hfill & {0} \hfill & {0} \hfill & {0} \hfill
\\
 {1} \hfill & {0} \hfill & {0} \hfill & {0} \hfill & {0} \hfill & {0} \hfill
\\
 {0} \hfill & {0} \hfill & {1} \hfill & {0} \hfill & {0} \hfill & {0} \hfill
\\
 {0} \hfill & {0} \hfill & {0} \hfill & {1} \hfill & {0} \hfill & {0} \hfill
\\
 {0} \hfill & {0} \hfill & {0} \hfill & {0} \hfill & {0} \hfill & {1} \hfill
\\
 {0} \hfill & {0} \hfill & {0} \hfill & {0} \hfill & {1} \hfill & {0} \hfill
\\
\end{array}}}  \right).
\end{equation}

The matrix of SWAP operation between qubits reads

\begin{equation}
SWAP = \left( {{\begin{array}{*{20}c}
 {1} \hfill & {0} \hfill & {0} \hfill & {0} \hfill & {0} \hfill & {0} \hfill
\\
 {0} \hfill & {0} \hfill & {0} \hfill & {0} \hfill & {1} \hfill & {0} \hfill
\\
 {0} \hfill & {0} \hfill & {0} \hfill & {1} \hfill & {0} \hfill & {0} \hfill
\\
 {0} \hfill & {0} \hfill & {1} \hfill & {0} \hfill & {0} \hfill & {0} \hfill
\\
 {0} \hfill & {1} \hfill & {0} \hfill & {0} \hfill & {0} \hfill & {0} \hfill
\\
 {0} \hfill & {0} \hfill & {0} \hfill & {0} \hfill & {0} \hfill & {1} \hfill
\\
\end{array}}}  \right).
\end{equation}

Fig. \ref{figure:SWAP} illustrates the plausible way to construct
SWAP. Indeed,

\begin{equation}
\hat {E} \cdot NOT_{1} \cdot NOT_{2} \cdot \hat {E} = \left(
{{\begin{array}{*{20}c}
 {1} \hfill & {0} \hfill & {0} \hfill & {0} \hfill & {0} \hfill & {0} \hfill
\\
 {0} \hfill & {0} \hfill & {0} \hfill & {0} \hfill & {1} \hfill & {0} \hfill
\\
 {0} \hfill & {0} \hfill & {0} \hfill & {1} \hfill & {0} \hfill & {0} \hfill
\\
 {0} \hfill & {0} \hfill & {1} \hfill & {0} \hfill & {0} \hfill & {0} \hfill
\\
 {0} \hfill & {1} \hfill & {0} \hfill & {0} \hfill & {0} \hfill & {0} \hfill
\\
 {0} \hfill & {0} \hfill & {0} \hfill & {0} \hfill & {0} \hfill & {1} \hfill
\\
\end{array}}}  \right) = SWAP.
\end{equation}

We should notice that matrices $NOT_{1}$ and $NOT_{2}$ commute,
i.e., $NOT_{1} \cdot NOT_{2} = NOT_{2} \cdot NOT_{1}$. Moreover,
$\hat {E} \cdot \hat {E} = \hat {E}^{2} = \hat {1}$.

Square root of NOT operation can be performed as NOT operation
with a half pulse duration. The matrices of such operations are

\begin{eqnarray}
\sqrt {NOT_{1}}  = {\frac{{1}}{{\sqrt {2i}}} }\left(
{{\begin{array}{*{20}c}
 {1} \hfill & {0} \hfill & {0} \hfill & {i} \hfill & {0} \hfill & {0} \hfill
\\
 {0} \hfill & {\sqrt {2i}}  \hfill & {0} \hfill & {0} \hfill & {0} \hfill &
{0} \hfill \\
 {0} \hfill & {0} \hfill & {1} \hfill & {0} \hfill & {0} \hfill & {i} \hfill
\\
 {i} \hfill & {0} \hfill & {0} \hfill & {1} \hfill & {0} \hfill & {0} \hfill
\\
 {0} \hfill & {0} \hfill & {0} \hfill & {0} \hfill & {\sqrt {2i}}  \hfill &
{0} \hfill \\
 {0} \hfill & {0} \hfill & {i} \hfill & {0} \hfill & {0} \hfill & {1} \hfill
\\
\end{array}}}  \right),\\
\sqrt {NOT_{2}}  = {\frac{{1}}{{\sqrt {2i}}} }\left(
{{\begin{array}{*{20}c}
 {1} \hfill & {0} \hfill & {i} \hfill & {0} \hfill & {0} \hfill & {0} \hfill
\\
 {0} \hfill & {\sqrt {2i}}  \hfill & {0} \hfill & {0} \hfill & {0} \hfill &
{0} \hfill \\
 {i} \hfill & {0} \hfill & {1} \hfill & {0} \hfill & {0} \hfill & {0} \hfill
\\
 {0} \hfill & {0} \hfill & {0} \hfill & {1} \hfill & {0} \hfill & {i} \hfill
\\
 {0} \hfill & {0} \hfill & {0} \hfill & {0} \hfill & {\sqrt {2i}}  \hfill &
{0} \hfill \\
 {0} \hfill & {0} \hfill & {0} \hfill & {i} \hfill & {0} \hfill & {1} \hfill
\\
\end{array}}}  \right).
\end{eqnarray}

\noindent The commutativity of matrices $\sqrt{NOT_{1}}$ and
$\sqrt{NOT_{2}}$ could be easily checked out. Then

\begin{eqnarray}
&& SWAP \nonumber\\
&& = \hat {E} \cdot NOT_{1} \cdot NOT_{2} \cdot \hat {E}
\nonumber\\
&& = \hat {E} \cdot \sqrt {NOT_{1}}  \sqrt {NOT_{1}} \cdot \sqrt
{NOT_{2}} \sqrt {NOT_{2}}
\cdot \hat {E} \nonumber \\
&& = \hat {E} \cdot \sqrt {NOT_{1}}  \sqrt {NOT_{2}}  \cdot \sqrt
{NOT_{1}} \sqrt {NOT_{2}}  \cdot \hat {E} \nonumber \\
&& = \hat {E} \cdot \sqrt {NOT_{1}}  \sqrt {NOT_{2}}  \cdot \hat
{E}\hat {E} \cdot \sqrt {NOT_{1}}  \sqrt {NOT_{2}}  \cdot \hat {E}
\nonumber \\
&& = \left( {\hat {E} \cdot \sqrt {NOT_{1}}  \sqrt {NOT_{2}} \cdot
\hat {E}} \right)^{2}.
\end{eqnarray}

\noindent Consequently,

\begin{eqnarray}
\sqrt {SWAP} && = \hat {E} \cdot \sqrt {NOT_{1}} \sqrt {NOT_{2}}
\cdot \hat {E} \nonumber\\
&& = {\frac{{1}}{{2i}}}\left( {{\begin{array}{*{20}c}
 {2i} \hfill & {0} \hfill & {0} \hfill & {0} \hfill & {0} \hfill & {0}
\hfill \\
 {0} \hfill & {1} \hfill & {i} \hfill & {i} \hfill & { - 1} \hfill & {0}
\hfill \\
 {0} \hfill & {i} \hfill & {1} \hfill & { - 1} \hfill & {i} \hfill & {0}
\hfill \\
 {0} \hfill & {i} \hfill & { - 1} \hfill & {1} \hfill & {i} \hfill & {0}
\hfill \\
 {0} \hfill & { - 1} \hfill & {i} \hfill & {i} \hfill & {1} \hfill & {0}
\hfill \\
 {0} \hfill & {0} \hfill & {0} \hfill & {0} \hfill & {0} \hfill & {2i}
\hfill \\
\end{array}}}  \right).
\end{eqnarray}

\begin{figure}
\includegraphics{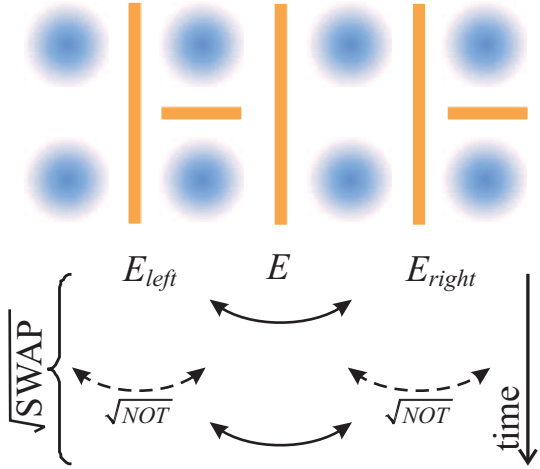}
\caption{\label{figure:Sqrt:SWAP} $\sqrt{SWAP}$ realization.}
\end{figure}

The scheme of square root of SWAP is shown in Fig.
\ref{figure:Sqrt:SWAP}. The dimensionality of all matrices is
$6\times 6$. This property leads to the particular form of the
operator of controlled phase shift

\begin{equation}
\hat {\Pi}  = \left( {{\begin{array}{*{20}c}
 {1} & {0} & {0} & {0} & {0} & {0} \\
 {0} & {1} & {0} & {0} & {0} & {0} \\
 {0} & {0} & {1} & {0} & {0} & {0} \\
 {0} & {0} & {0} & {1} & {0} & {0} \\
 {0} & {0} & {0} & {0} & {1} & {0} \\
 {0} & {0} & {0} & {0} & {0} & { - 1} \\
\end{array}}}  \right).
\end{equation}

As far as matrices $4\times 4$ are concerned, XOR operation is

\begin{eqnarray}
\left( {\hat {Z}_{1} (\pi/2) \otimes \hat {Z}_{2} (-\pi/2) }
\right) \cdot \sqrt {SWAP} \nonumber \\ \times \left( {\hat
{Z}_{1} (\pi ) \otimes \hat {1}_{2}} \right) \cdot \sqrt {SWAP},
\end{eqnarray}

\noindent where $\hat{Z}$ is the phase shift gate \cite{loss98}.
In a similar way, a direct calculation shows that in our case

\begin{eqnarray}
\hat {\Pi} = && \left[ {\left( {\hat {Z}_{1} (\pi/2) \otimes \hat
{Z}_{2} ( - \pi/2)} \right) \cdot \sqrt {SWAP}} \right]^{2}
\nonumber\\ && \times \left( {\hat {Z}_{1} (\pi ) \otimes \hat
{1}_{2}} \right) \cdot \sqrt {SWAP}.
\end{eqnarray}

\noindent Eventually, CNOT operation looks like

\begin{equation}
CNOT = \left( {\hat {1}_{1} \otimes \hat {H}_{2}}  \right) \cdot
\hat {\Pi}  \cdot \left( {\hat {1}_{1} \otimes \hat {H}_{2}}
\right),
\end{equation}

\noindent where $\hat{H}$ is Hadamard's transformation:

\begin{equation}
\hat {1}_{1} \otimes \hat {H}_{2} = {\frac{{1}}{{\sqrt {2}}}
}\left( {{\begin{array}{*{20}c}
 {1} & {0} & {1} & {0} & {0} & {0} \\
 {0} & {\sqrt {2}}  & {0} & {0} & {0} & {0} \\
 {1} & {0} & { - 1} & {0} & {0} & {0} \\
 {0} & {0} & {0} & {1} & {0} & {1} \\
 {0} & {0} & {0} & {0} & {\sqrt {2}}  & {0} \\
 {0} & {0} & {0} & {1} & {0} & { - 1} \\
\end{array}}}  \right).
\end{equation}

\newpage

\bibliography{references_06_03_09}

\end{document}